# Controlling magnetoresistance by oxygen impurities in Mq3-based molecular spin valves

*Alberto Riminucci\*†, Zhi-Gang Yu‡, Mirko Prezioso†∥, Raimondo Cecchini†⊥, Ilaria Bergenti†, Patrizio Graziosi†#, Valentin Alek Dediu†*

†CNR-ISMN, via P.Gobetti 101, 40129 Bologna, Italy

‡ISP/Applied Sciences Laboratory, Washington State University, Spokane, Washington 99210, USA

**Abstract**

The understanding of magnetoresistance (MR) in organic spin valves (OSVs) based on molecular semiconductors is still incomplete after its demonstration more than a decade ago. While carrier concentration may play an essential role in spin transport in these devices, direct experimental evidence of its importance is lacking. We probed the role of charge carrier concentration by studying the interplay between MR and multilevel resistive switching in OSVs. The present work demonstrates that all salient features of these devices, particularly the intimate correlation between MR and resistance, can be accounted for by the impurity band model, based on oxygen migration. Finally, we highlight the critical importance of carrier concentration in determining spin transport and MR in OSVs and the role of interface-mediated oxygen migration in controlling the OSVs response.

# 1. INTRODUCTION

Spin injection and transport in non-magnetic media has given rise to a wealth of novel physics [1] and to sustained technological innovations [2]. While spintronics phenomena were initially observed in inorganic materials, increasing research efforts were devoted to organic semiconductors (OS) due to their intrinsically low spin-orbit coupling and hyperfine interactions[3].

One of the few unambiguous indications of spin injection in non-magnetic conductive media is the presence of spin precession or, equivalently, of the Hanle effect[4]. In organic pseudo spin valves (OSVs), spin injection and transport were initially studied through magnetoresistance (MR), that is the dependence of the electrical resistance of an OSV on the relative orientation of the magnetization of its ferromagnetic electrodes. The fact that the effect of spin precession was not observed [2,5] (despite the presence of electron paramagnetic resonance which is due to magnetic-field induced spin precession[6]), prompted the introduction of a new model of spin transport in OSs[7] which is controlled by the charge carrier density. This was later expanded by introducing a microscopic mechanism, based on doping impurities, to describe charge and spin transport in OSVs [8]. This impurity band model draws a clear connection between the charge transport and the spin transport properties of an OS. In this model the density of carriers not only dictates the conductive properties but also the spin transport ones, via an exchange mechanism between carriers.

$La_{0.7}Sr_{0.3}MnO_3$(LSMO)/tris-(8-hydroxyquinoline)metal(Mq3)/AlOx/Co OSVs, where Mq3 is an organic molecular semiconductor and M can be either Aluminium or Gallium, are an ideal playground to validate the impurity band model. In these devices both carrier density and magnetoresistance can be modulated systematically through multilevel resistive switching[9–13]. The study of multilevel resistive switching in OSs is motivated primarily by their considerable applicative potential [12,14–17]. While several models for multilevel resistive switching were put forward for organic materials[15–17], none of them involved ferromagnetic electrodes and spin transport and therefore cannot describe OSVs. Conformational changes of Alq3 were proposed to explain resistive switching in Alq3 based devices[18], but the current voltage characteristics, according to that model, would follow an exponential behaviour, in contrast to the ones observed in our experiments. A common mechanism for resistive switching is filamentary conduction due to the existence of preferential conductive paths on the Alq3 layer. We exclude this mechanism in its simplest forms of metallic protrusions[19] or carbon rich paths[20] since our current-voltage (I-V) characteristics are non-ohmic. The possible inclusion of Co filaments in our devices was investigated in prior work[21], were it was demonstrated that the AlOx layer acted as an effective barrier preventing Co inter-diffusion.

Interestingly, theoretical calculations predict highly conductive paths that are intrinsic to Alq3 and not generated by metallic inclusions[22]. In these filaments both mobility and carrier density can be higher than in the rest of the OS. A more complex filamentary conduction mechanism involves the existence of highly conductive paths due to local doping by some impurity such as $O_2$, which acts as an acceptor. The presence of $O_2$ is known to greatly affect charge transport in Alq3[23] and $O_2$ doping effects become important above an exposure of ~$5\times10^3$ mbar s[24].

In this article we show that both resistance and magnetoresistance in LSMO/tris-(8-hydroxyquinoline)aluminum(Alq3)/AlOx/Co or LSMO/tris-(8-hydroxyquinoline)gallium(Gaq3)/AlOx/Co OSVs exhibit multilevel resistive switching, which can be consistently accounted for by the impurity band model. Based on recent spectroscopic results[25] we propose a mechanism based on $O_2$ doping of the molecular layer, due to the migration of $O_2$ from the AlOx barrier into the OS and *vice versa*. We measured the magnetoresistance of the devices in several resistive states, for which we provided an electrical characterization. We also investigated the temperature dependence of the electrical and of the magnetoresistive properties. The model was able to reproduce all the salient features satisfactorily, thus providing a comprehensive understanding of charge and spin transport phenomena in OSVs.

**2. EXPERIMENTAL SECTION**

We have studied the behaviour of LSMO/Alq3/AlOx/Co or LSMO/Gaq3/AlOx/Co OSVs with a molecular film thicknesses varying between 10 nm and 300 nm. The LSMO was grown by channel spark ablation at an $O_2$ pressure of $10^{-2}$ mbar on a matching $SrTiO_3$ substrate[26]. The LSMO was then placed in the organics deposition chamber, with a base pressure of $10^{-8}$ mbar. The LSMO was annealed for 30' at 250 °C to clean its surface after exposure to ambient atmosphere. Without breaking the vacuum, Gaq3 or Alq3 was sublimated from a Knudsen cell, whose temperature was 261 °C. The deposition rate was 0.13 Å/s. In the same vacuum chamber, a 2.4 nm thick Al film was deposited on top of the molecular layer at a rate of 0.2 Å/s with a cell temperature of 473 °C. The Al film was then oxidized for 15' in a $10^2$ mbar atmosphere of pure $O_2$ (total exposure $9\times10^4$ mbar s). The sample was moved to the metal deposition chamber with a base pressure $<5\times10^{-9}$ mbar, after changing the shadow mask in ambient air. Finally, the top Co 20 nm thick layer was deposited at a rate of 0.2 Å/s. The devices' active area was $0.2\times1 mm^2$.

The top Co electrode was connected to the ground and the bias voltage was connected to the bottom LSMO one. The magnetoresistance was defined as MR= $(R_{AP}-R_P)/R_P$, where $R_P$ is the resistance of the OSV when the magnetization of the electrodes are parallel to each other and $R_{AP}$ is the resistance of the OSV when the electrodes are oriented antiparallel to each other. The MR was measured at an applied bias voltage of -0.1 V.

The resistance measurements in Figures 1 and 3 were taken by applying a "reading" bias voltage of -0.1 V, which was small enough no to perturb the resistive state. The resistance changes were obtained by applying a "writing" bias voltage, high enough to affect the resistive state. Typically this voltage was between 2 V and 4 V in absolute value. The application of a positive "writing" voltage caused the resistance to decrease, while the application of a negative one cause the resistance to increase.

**3.RESULTS AND DISCUSSION**

The OSVs devices studied in this work are schematically depicted in Figure 1a). Figure 1b) provides an overview of the results on the resistance as a function of the thickness of the molecular layer, obtained in our laboratory and from the literature; the active area was normalized to 1mm$^2$. While differences in behaviour between Gap3 and Alq3 were reported in the literature[27], they are negligible compared to OSVs device-to-device variability, which is a common occurrence for memristive devices[28]. It is well-established from theoretical calculations[29] that in metal quinoline complexes, the highest occupied molecular orbital (HOMO) is located on the phenoxide side of the quinolinol ligands, whereas the lowest unoccupied molecular orbital (LUMO) is found on the pyridyl side; the central metal ion contributes negligibly to the density of states. The marginal role of the central metal ion in the spin polarized effects was predicted by demonstrating that the electronic and magnetic properties of the interface between Mq3 molecules and Co could not be modified by chemically substituting the metal ion[30]. Finally, magnetoresistance and multilevel resistive swiching were observed in both Gaq3[25] and Alq3[10] based devices.

In our experiments no difference in behaviour between Alq3 and Gaq3-based OSVs was detected, therefore in this article we will use their results indifferently. In Figure 1b), red triangles represent data from devices made in our laboratory that showed MR. Blue circles refer to devices from the literature; more details about these data can be found in Table S1 in the Supporting Information. Black squares represent data from devices made in our laboratory that did not show MR and which typically have resistances greater than 5 MΩ[31]. The resistance of the devices made in our laboratory and of those from the literature did not show a clear dependence on the thickness of the molecular semiconductor layer. This suggests that the device resistance was controlled by the electrode/OS interfaces, i.e. charge transport was injection limited.

The devices fabricated in our laboratory showed multilevel resistive switching[13] and also an interplay between the resistive state and the MR[10](see Supporting information, Figure S1). As we will show, these observations can be explained with the impurity band model[8] by taking $O_2$ as the impurity; in our devices $O_2$ acted as an acceptor, therefore the carriers' concentration was proportional to the number of $O_2$ impurities present in the OS. The details of the model are reported in the section "Device Resistance" of the Supporting Information.

In the impurity band model, the resistance of the device originates mainly from the interface between the electrodes and the molecular layer. The multilevel resistive switching observed in our devices can be explained by the ability to control the $O_2$ impurity concentration by the application of a voltage pulse[25]. In the high resistance state, the available $O_2$ is concentrated in the AlOx barrier while in the low resistance state it partially migrates into the molecular layer. In this case, $O_2$ acts as an acceptor and its migration from the AlOx into the molecular layer causes the concentration of hole carriers in the OS to increase. In turn, the carriers formed by $O_2$ allow charge storage at the interface, with an decrease in the injection barrier thickness[13,32] and thus of the resistance. In addition, an increased number of vacancies in the AlOx, makes the tunnel barrier leaky, which increases its conductivity and improves its spin injection properties: a leaky barrier allows carriers to be injected into the molecular layer while a non-leaky one acts simply as a tunnelling barrier, jointly with the molecular layer[33].

A sufficiently large voltage of opposite polarity causes the $O_2$ to migrate back into the AlOx barrier. A decrease in the number of carriers in the molecular layer reduces charge storage at the interface with a concomitant increase of the device resistance due to the increased thickness of the depletion region. At the same time, a greater amount of $O_2$ in the AlOx layer makes the tunnel barrier less leaky, thus increasing its resistance.

It must be stressed that central ion in metal quinolines plays no role in $O_2$ migration. Experimental results on the Carbon spectra clearly indicate that the effects of $O_2$ migration involve only the HOMO levels and leave the metal ion spectra unaffected [25]. $O_2$ diffusion is the associated with a partial redistribution of the electronic charge within the complex, producing a positively charged state the quinoline ligand.

Models of conduction based on the existence of an impurity band in the gap of an insulator can be traced back to the work by Hickmott et al.[34]. In the intervening time, much work was carried out on both organic and inorganic insulators[35]. The existence of the impurity band was attributed to the presence oxygen vacancies[36] or metallic ions[37] in the insulating layer. In all cases, the impurity band was associated with resistive switching, as in the present work.

The role of oxygen in many instances of multilevel resistive switching is well established in inorganic devices[36,38]. Resistive switching was reported in LSMO itself[39], in which an electrical contact was made with a conductive atomic force microscopy tip. However, direct metallic contact between the electrodes in our devices is ruled out as most of the voltage drop occurs either across the molecular layer or across the tunnelling barrier, as we demonstrate in what follows. The resistivity of the LSMO films in the present work is approximately $10^{-5}$-$10^{-6}$ $\Omega$m[40]. This means that in our devices LSMO cannot support an electrical field intense enough to cause internal oxygen migration. Such hypothesis was put forward by Grünewald et al. in devices with a much lower resistance than in our devices (10s of k$\Omega$ for a smaller, 0.2×0.15 mm$^2$ active area, which translates in 100s of $\Omega$ when scaled to 1×1 mm$^2$), hence with a greater ability to sustain an electric field in the LSMO [11]. We therefore conclude that in our devices we cannot provide electric fields strong enough to affect the LSMO electrode.

In general, resistive switching can be classed as unipolar or bipolar. In the former case, the high and the low resistance states are reached by applying bias voltages of the same sign, while the latter one, voltages needed to increase and decrease the resistance have opposite polarities. While unipolar switching is associated with diffusion or temperature phenomena, bipolar switching is driven by the electric field[41]. In the present devices, resistive switching could be taking place solely in the AlOx tunnel barrier[36]. An interplay between conductive paths in AlOx, due to oxygen vacancy formation, and corresponding electrons trapped in the OS were observed in AlOx/polymer sandwich devices[42]. However, they observed unipolar switching, in contrast to the bipolar switching we observed. Similar conductive paths were observed on Alq3 based devices[43].

Let's now tackle the issue of magnetoresistance in our devices. While spin transport has been mostly elucidated in inorganic metals and semiconductors [44], the understanding of spin injection and transport in molecular materials and devices is still primitive. Generally speaking, the spin scattering mechanisms that work for inorganic, ordered materials[45,46] are not directly applicable to disordered molecular semiconductors. Instead, spin-dependent hopping due to spin-orbit coupling[27,47] and hyperfine interaction[48] are believed to be the two main spin scattering mechanisms in OSs.

At lower temperatures than the ones used in the present work, namely at 4.2 K, the observed MR was attributed to tunnelling anisotropic magnetoresistance (TAMR) from LSMO into Alq3[5]. The same mechanism was used to explain the change in MR that went together with resistive switching[11] which was in turn was attributed to oxygen migration in the LSMO. The working temperature for the present results is 100 K or greater, and it exceeds the maximum temperature at which TAMR can be observed in LSMO.

The main obstacle to spin injection in inorganic semiconductors is the conductivity mismatch between the low resistance electrodes and the high resistance inorganic layer[49], and can be circumvented by introducing a tunnelling barrier between the injecting electrode and the inorganic semiconductor[50,51]. On the contrary, conductivity mismatch does not play a great role in OSVs [8,52]. Numerous works were devoted to the demonstration of spin injection in an organic semiconductor. Results from ferromagnetic/organic semiconductors/ferromagnetic OSVs often reported MR [33,53,54], but this alone is not considered to be a definitive demonstration of spin injection. Two photon photo emission spectroscopy was used by Chinchetti et al. to study spin injection in CuPc, but did not probe electrical spin transport[55]. Drew et al. used muon spin rotation to detect spin polarization in a Alq3/N,N'-Bis(3-methylphenyl)-N,N'-diphenylbenzidine(TPD) device[56] but the interpretation of experimental data was very complex and highly model dependent. Nguyen et al. studied the isotope effect by using deuterated

poly(dioctyloxy)phenylenevinylene (DOO-PPV) and demonstrated that the main spin scattering mechanism was due to the hyperfine field[48]. The isotope effect, however, was found to be absent in Alq3 [57], where Nuccio et al. demonstrated spin orbit coupling to be the main spin scattering mechanism[27]. Watanabe et al. demonstrated spin pumping across poly(2,5-bis(3-alkylthiophen- 2-yl)thieno[3,2-b]thiophene) (PBTTT) [58] and Jiang et al. did in Alq3[59]. In spin pumping no net charge current flows between the electrodes and carriers are produced by unintentional doping, in analogy with the impurity band model.

Within the framework of the impurity band model, spin transport and thus magnetoresistance in our devices are strongly related to the $O_2$ impurity concentration and hence to the resistive state of the device. The $O_2$, acting an acceptor dopant in the OS, introduces mobile polarons that act as charge carriers. In disordered organic semiconductors, these polarons are spatially localized. This localization suggests an exchange interaction between the spins of adjacent polarons, $H_{ex} = 2\bar{J}\, S_1 \cdot S_2$, where the exchange integral $\bar{J}$ can be estimated from[60]:

$$\bar{J} = 0.821 \frac{e^2}{\epsilon\, \xi} \left(\frac{\bar{R}}{\xi}\right)^{5/2} e^{-2\bar{R}/\xi} \quad (1)$$

where $\epsilon$ is the dielectric constant, $\xi$ is the electron localization length, and $\bar{R}$ is the average distance between polarons. The exchange coupling can cause an effective spin motion, giving rise to an additional spin diffusion constant, $D \sim \bar{J}\, R^2$ [12]. Hence high-efficiency spin transport is achieved via the exchange interaction when a sufficiently high carrier concentration is reached, since according to equation (1), $\bar{J}$ is exponentially dependent on the carriers' concentration via $\bar{R} = 1/n^3$. Crucially, the exchange mechanism contained in the impurity band model allows spin transport to take place without an accompanying charge transport, i.e., spin transport is decoupled from charge transport. The high spin transport velocity of this mechanism explains the absence of the Hanle effect.

Finally, the impurity band model includes spin-polarized tunnelling at the two electrodes. The coefficients for current injection from the ferromagnetic electrodes ($A_1$ and $A_2$ in the "Device resistance" section of the Supporting Information) become spin-polarized. Spin diffusion inside the molecules contains contributions from both hopping, which contributes also to charge current, and from exchange between neighbouring electrons, which does not contribute to charge current. It should be noted that while the charge current is continuous over the device, the spin current is not, because of possible spin relaxation at the interface. The detailed formalism can be found in Ref. [17]. The device resistance was calculated by using the tunnelling expressions in equations (3) and (4) in Yu[8].

The central role of the carrier's concentration was confirmed by the fact that in low carrier concentration OSVs, neither magnetoresistance nor multilevel resistive switching were observed[31]. Instead in the devices at hand, carrier concentration was sufficiently high to establish the exchange interaction.

The above picture of multilevel resistive switching and magnetoresistance was used to model the data in Figure 1c) and Figure 1d). To obtain these data, the resistance of the OSV was changed by a suitable voltage pulse and subsequently measured at -100 mV. Figure 1c) shows how the resistive state affects the MR of the devices. The black dots are data taken from a 200 nm thick device at 100 K, while the black line is a fitting with the impurity band model. Since the carriers induced by acceptor $O_2$ are of p-type, the barrier height between a metal and the organic is $q\phi_B = E_g - q(\phi_m - \chi)$, where $E_g$ is the gap between LUMO and the HOMO, and $q\chi$ the electron affinity. The barrier heights were set to 1.5 eV for LSMO and to 1.7 eV for Co, which are consistent with the values measured via photoelectron spectroscopy, 1.7 eV and 2.1 eV, in a similar device [61,62]. Figure 1d) shows the MR and resistance as a function of the carrier concentration. The black dots represent experimental data on the resistance, while the red ones represent the corresponding MR. The solid lines represent a fit with the model. Clearly, the carrier concentration captures very well the relationship between the resistance and the MR. This validates the carrier concentration as the crucial parameter determining the MR and the resistance of OSVs.

Figures 2a) and 2b) show a schematic description of the energy levels of the different layers making up the OSVs. They show the band alignments under zero bias for high impurity concentration (Figure 2a) and low impurity concentrations (Figure 2b). At zero bias, the device is in equilibrium with a common Fermi level throughout the entire molecular layer. Under a finite bias, according to the impurity model and our simulation results, the voltage drop occurs mainly at the two depletion regions and the electrochemical potential (quasi-Fermi level) in the organic interior remains fairly flat[13].

The impurity band is located between the black dotted lines, with the impurities being represented by the red dashes. $V_{b1}$ and $V_{b2}$ are the built-in voltages. Figure 2a) shows a state with a high $O_2$ impurity concentration in the OS. This is associated with an increased number of oxygen vacancies (white circles) in the AlOx barrier (grey vertical bar). $d_{H1}$ and $d_{H2}$ are the Schottky barrier thicknesses in this case. Figure 2b) shows the case of a lower concentration of $O_2$ impurities, with a correspondingly smaller number of oxygen vacancies in the AlOx barrier. In this case the Schottky barriers' thicknesses are $d_{L1}$ and $d_{L2}$ for the two electrodes respectively. Since more charge is available when the impurity concentration is higher, the screening is more effective and hence $d_{H1}<d_{L1}$ and $d_{H2}<d_{L2}$. The Schottky barriers' thicknesses, the carrier concentration in the OS and the concentration of oxygen vacancies in the AlOx barrier determine the overall resistance of the OSV

The importance of carrier concentration and resistance to MR in OSVs is further corroborated by impedance spectroscopy on the low and the high resistance states at 100 K. The data are shown in Figures 2c) ad 2d) respectively. The data were modelled with a simple electrical circuit in which a resistance $R_S$ is connected in series with an RC loop, as shown in the inset to Figure 2c); $R_S$ is the resistance coming from electrodes and interfaces, R is the resistance coming from the device interior, and C is the device capacitance. By fitting the circuit model we find that for the low-resistance state, $R_s$ =(428±5) Ω and R=(1339±5) Ω, whereas for the high-resistance device, $R_s$ =(500±27) Ω and R=(14.1±1.4) MΩ. The capacitance for the low- and high-resistance devices was (230±1) pF and (217.2±0.1) pF, respectively, which are virtually identical and close to the geometrical capacitance of 133 pF.

The small interior resistance R in the low-resistance device indicates a large carrier concentration in the bulk region. By using the device resistance and the device dimensions to estimate the carrier density n, one obtains values as high as n=$10^{20}$ cm$^{-3}$ if the Alq3 mobility is used[63]. Moreover, in the low-resistance device, the interfacial and interior resistances are comparable, whereas in the high-resistance device, the interior resistance is orders-of-magnitude larger than the interfacial resistance. The lack of MR in the high-resistance device can also be understood in terms of the severe conductivity mismatch between the interfacial and the interior resistance. The low-resistance device, on the other hand, does not suffer from such resistance mismatch and exhibits a pronounced MR.

To gain a deeper understanding of the transport mechanism in the OSVs, we studied their electrical resistance as a function of temperature. In the high resistance state the resistance decreases monotonically as the temperature is increased (see Supporting Information, Figure S2), as expected for a semiconductor material. Figure 3a) shows the resistance as function of temperature for a similar device after it was set in the low resistance state; the black circles indicate the experimental data points measured at -100 mV, while the red solid line is the model fit. The behaviour in this case was non-monotonic: the resistance increased as the temperature was increased from 100 K, it reached a maximum and then it decreased for higher temperatures. To model the temperature dependence of the resistance, we have considered two processes. First, the resistance of the LSMO electrode increases with temperature in its ferromagnetic phase because of its change in magnetization, and consequently the resistance due to tunnelling from the LSMO electrode increases with temperature. Second, the conductivity in the OS region decreases with temperature in the low-temperature regime, which is similar to metallic conduction and can be considered in terms of scattering. At high temperatures, carriers can use thermally activated hopping for conduction, and the conductivity increases with temperature. In the impurity band model, the crossover temperature for the data in Figure 3a) was set to 270 K and the activation energy was 0.2 eV, compatible with the values that can be found in the literature[64]. There is clearly a good agreement between the model and the experimental data.

A similar dependence of the resistance on temperature was observed in tunnelling inorganic devices with LSMO electrodes and was attributed to oxygen deficiency at the interface between LSMO and the tunnelling layer [65–67]. If this were the case here, the application of a magnetic field would lead to an increase of the temperature of the resistance maximum. We carried out this measurement at 80 mT and no shift was observed (see Supporting Information, Figure S3). A non monotonic behaviour of the resistance as a function of temperature was observed in some doped polymers[68] but it was opposite to ours, that is the polymeric organic semiconductor showed a resistance minimum and not a maximum, as a function of temperature.

The different transport behaviours at low and high temperatures also manifest themselves in the I-V characteristics of the devices at different temperatures. In this case the information comes from the dependence of the differential conductance on the applied voltage. There was a sharp distinction between the differential conductance observed at 150 K and that observed at 300 K. At 300 K the first derivative of the differential conductance curve was continuous at 0 V bias (Figure 3b), blue dots) whereas at lower temperatures, the differential conductance at 0 V has a kink (Figure 3b), black dots); the model fits are given by the respective solid lines. The difference in behaviour is due to the different electric field dependence in the high and low temperature regimes. At low temperatures, the carriers in the bulk region use band conduction, or tunnelling between adjacent sites of the impurity band. In this situation, the tunnelling probability increases with an electric field. One can think that the tunnelling barrier is reduced by the electric field. By contrast, at high temperatures, the carriers use hopping for conduction. An electric field makes the energy at different sites different, and therefore the carrier mobility actually decreases with the electric field, when the field is not too strong.

**4.CONCLUSIONS**

We have investigated the behaviour of several Gaq3 and Alq3 –based OSVs that consistently showed multilevel resistive switching and magnetoresistance. All the salient features of these devices can be consistently explained with the impurity band model. In this model, $O_2$ acted as acceptor dopant of the OS and its concentration, dominated by interfacial migration, dictated both the conductive and the spintronic properties of the OSVs. Crucially, we demonstrated and elucidated the dependence of the MR on the resistive state of a device by the $O_2$ concentration in the OS. Our work provides a comprehensive understanding of the multilevel resistive switching and MR in OSVs as well as their intimate correlation.

■ ASSOCIATED CONTENT

**Supporting Information**

Table surveying the literature, relationship between the resistive state and the magnetoresistance, resistance as function of temperature and theoretical fit, resistance as function of temperature with and without an applied magnetic field, device resistance, LSMO electrode resistance and magnetoresistance as a function of temperature for 3 different devices, fit of the magnetoresistance of a device as a function of its resistance, resistance and corresponding magnetoresistance as a function of temperature, magnetoresistance as function of resistance when the temperature is varied, device resistance theoretical calculations

■ AUTHOR INFORMATION

**Corresponding Author**

*E-mail: a.riminucci@bo.ismn.cnr.it. Phone: +390516398509. Fax: +390516398540

**ORCID**

Alberto Riminucci: 0000-0003-0976-1810

Zhi-Gang Yu: 0000-0002-1376-9025

Mirko Prezioso: 0000-0002-3318-9132

Raimondo Cecchini: 0000-0003-3650-2478

Ilaria Bergenti: 0000-0003-0628-9047

Patrizio Graziosi: 0000-0003-0568-0255

Alek Dediu: 0000-0001-9567-5112

**Present address**

∥ Department of Electrical and Computer Engineering 9560, University of California Santa Barbara, Santa Barbara, California 93106, USA

⊥ CNR-IMM, via C. Olivetti 2, 20864 Agrate Brianza, (MB), Italy

# School of Engineering, University of Warwick, Coventry CV4 7AL, United Kingdom

**Author Contributions**

The manuscript was written through contributions of all authors. All authors have given approval to the final version of the manuscript.

**Funding**

This work is funded through the European Union Seventh Framework Programme (FP7/2007-2013) project Organic-inorganic hybrids for electronics and photonics(HINTS), under grant agreement GA no. 263104, and Italian Grant PRIN QCNaMos.

**Acknowledgements**

The authors would like to acknowledge the invaluable technical help received from Mr.Federico Bona.

**Figure 1**: a) Schematic diagram of the devices studied. The bottom electrode is made of $La_{0.7}Sr_{0.3}MnO_3$ on which an Alq3/AlOx/Co or Gaq3/AlOx/Co stack was deposited. OS thickness varied between 10 nm and 300 nm. b) Resistance as function of OS thickness for devices with different behaviours. The red triangles represent data from the devices studied in our laboratory that showed MR. For comparison, the blue circles represent data obtained from the literature (see Supporting Information, Table SI). Finally the black squares show data from nominally identical devices fabricated in our laboratory that did not show MR; these were the subject of a separate study[31]. c) MR as a function of the resistance of a LSMO/Alq3(200 nm)/AlOx/Co device, measured at 100 K, - 100 mV. The resistive states were set by the application of a suitable voltage pulse. The black circles represent experimental data while the solid line represents the model fit. d) The MR and R can be both calculated from the carrier concentration. From the same concentration value, both MR and R can be accurately computed with our model, and the MR vs R can be reproduced by sweeping the value of the impurity concentration (solid lines).

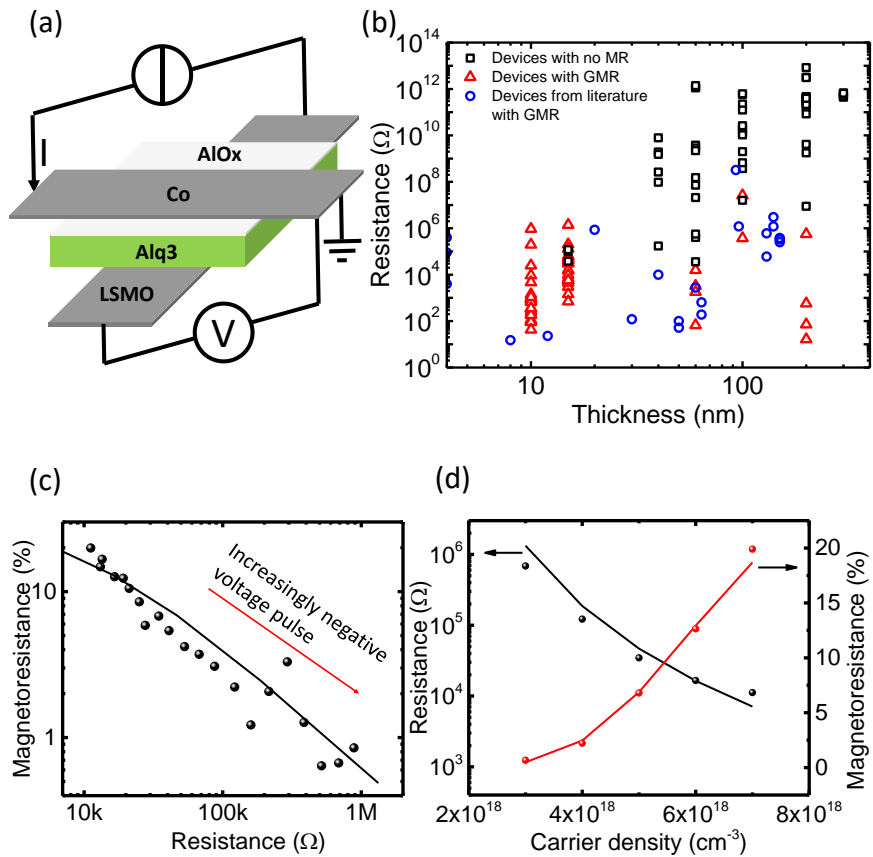

**Figure 2**: Panels a) and b) depict the energy level alignment of the composing elements of the devices. The black rectangles represent the electrodes while the grey ones represent the AlOx tunnel barriers. The dash-dotted line indicates the Fermi level, the dotted lines enclose the impurity band, the black solid lines represent the highest occupied molecular level (HOMO) while the red dashes represent the impurities. $V_{b1}$ and $V_{b2}$ are the built in voltages at the Schottky barriers. a) $d_{H1}$ and $d_{H2}$ are the Schottky barriers' thicknesses for the two electrodes in the case of high impurity concentration. b) $d_{L1}$ and $d_{L2}$ are the Schottky barriers' thicknesses in the case of low impurity concentration. Since the available charge is smaller in the latter case, the screening is less effective, therefore $d_{H1} < d_{L1}$ and $d_{H2} < d_{L2}$. The resistance of the device can be divided up in the Schottky barriers' one ($R_S$) and the interior one (R) due to the OS and the AlOx. A higher impurity concentration in the OS corresponds to a smaller resistance.

Panels c) and d) show the data obtained from impedance spectroscopy carried out on a LSMO/Alq3(200 nm)/AlOx/Co device at 100 K. The response was modelled by the circuit depicted in the inset. The geometrical capacitance was $\varepsilon \cdot \varepsilon_r \cdot A/d = 8.85 \times 10^{-12} \times 3 \times 10^{-6}/(2 \times 10^{-7}) = 133$ pF, with $\varepsilon = 3^{69}$. c) Impedance spectroscopy in the low resistance state. The series resistance, due to the contribution of the Schottky barriers, was $R_S = (428 \pm 5)$ Ω. The interior resistance was $R = (1339 \pm 5)$ Ω while the capacitance was C= $(230 \pm 1)$ pF. d) Impedance spectroscopy in the high resistance state. $R_S$ increased to $R_S = (500 \pm 27)$ Ω. The interior resistance increased to $R = (14.1 \pm 1.4)$ MΩ, while the capacitance was left virtually unvaried at C= $(217.2 \pm 0.1)$ pF.

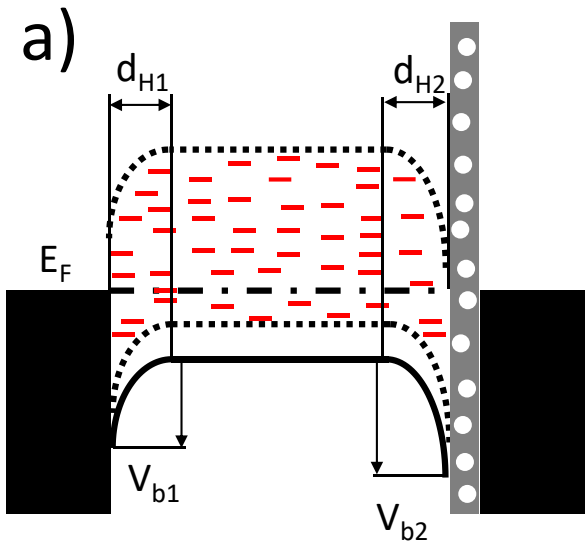
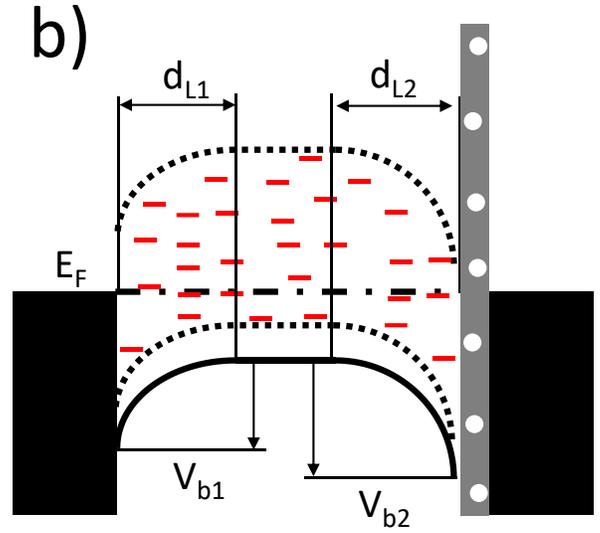
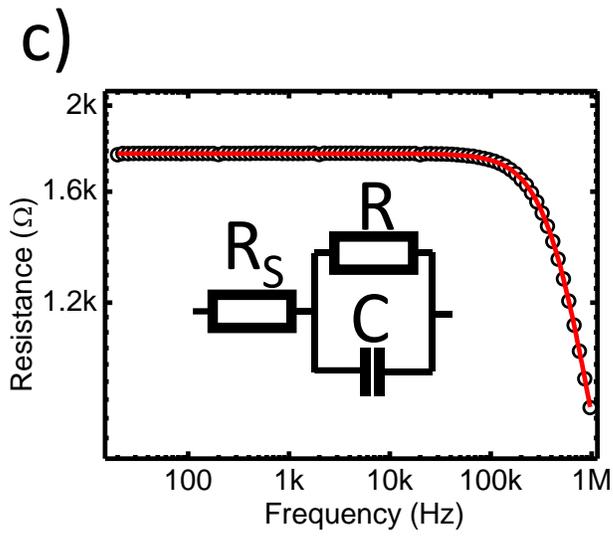
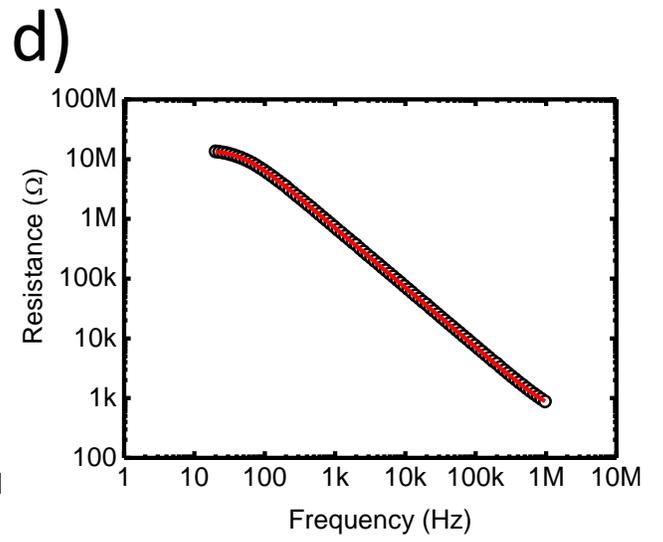

**Figure 3**: a) Resistance as function of the temperature for a LSMO/Gaq3(15 nm)/AlOx/Co device in the low resistance state. The black circles represent the measure data, while the red line represents the fit with the model. The behaviour was non-monotonic and was modelled by considering band transport in the OS at lower temperatures and hopping transport in the OS above the resistance hump's temperature. b) Differential conductance plots for a 15 nm thick Gaq3 device. The black circles correspond to data taken at 150 K, while the blue ones refer to data taken at 300 K. There is a sharp distinction between the two regimes, with a more abrupt change of slope at 0 V for the data at 150 K. The solid lines of the corresponding colour are fits with the impurity band model, and reproduce remarkably well the 0 V slope of the differential conductance at 150 K and 300 K.

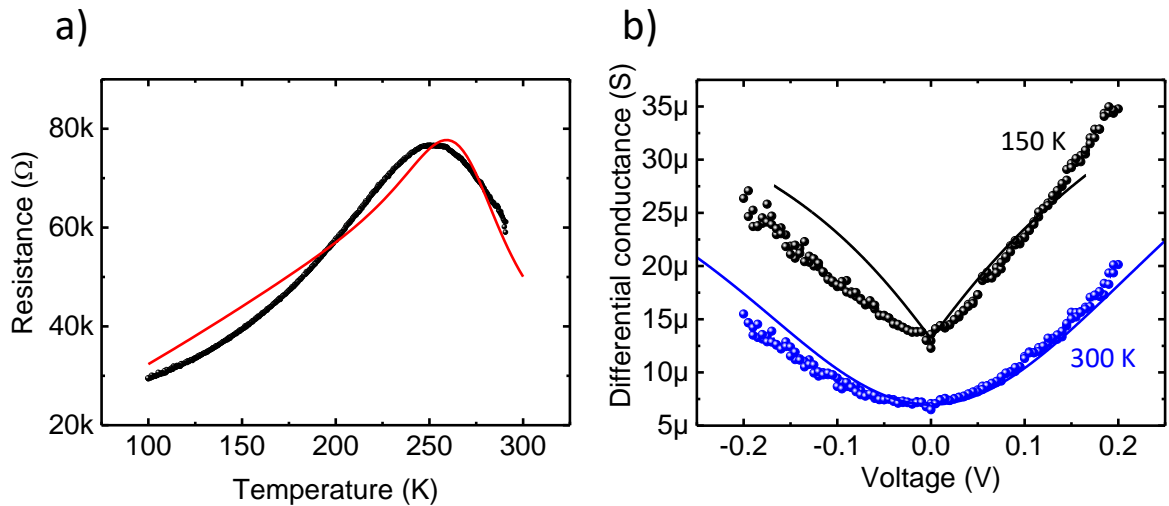